\makeatletter

\newenvironment{inlinefigure}{%
\def\@captype{figure}%
\noindent\begin{minipage}{0.999\linewidth}\begin{center}}
{\end{center}\end{minipage}\smallskip}
\makeatother

\documentclass[10pt,preprint]{aastex}
\usepackage{emulateapj5,epsfig}

\newcommand{\mdot}{\dot{M}}

\newcommand{\msun}{{M}_{\odot}}
\newcommand{\rsun}{{R}_{\odot}}
\newcommand{\msyr}{\msun \ {\rm yr^{-1}}}

\slugcomment{To appear in the Astrophysical Journal}

\shorttitle{Flickering in CH Cygni}
\shortauthors{Sokoloski \& Kenyon}

\begin{document}


\title{CH Cygni II: Optical Flickering from an Unstable Disk}


\author{J. L. Sokoloski and S. J. Kenyon}
\affil{Smithsonian Astrophysical Observatory, 60 Garden St., Cambridge, MA 02138}
\email{jsokoloski@cfa.harvard.edu}


\begin{abstract}
CH Cygni began producing rapid, stochastic optical variations with the
onset of symbiotic activity in 1963.  We use changes in this
flickering between 1997 and 2000 to diagnose the state of the
accretion disk during this time.  In the 1998 high state, the
luminosity of the $B$-band flickering component was typically more
than 20 times higher than in the 1997 and 2000 low states.  Therefore,
the physical process or region that produces the flickering was also
primarily responsible for the large optical flux increase in the 1998
high state. Assuming that the rapid, stochastic optical variations in
CH Cygni come from the accretion disk, as in cataclysmic variable
stars, a change in the accretion rate through the disk led to the 1998
bright state.  All flickering disappeared in 1999, when the accreting
WD was eclipsed by the red giant orbiting with a period of
approximately 14 yr, according to the ephemeris of Hinkle et al.  and
the interpretation of Eyres et al.  We did not find any evidence for
periodic or quasi-periodic oscillations in the optical emission from
CH Cygni in either the high or low state, and we discuss the
implications for magnetic propeller models of this system.  As one
alternative to propeller models, we propose that the activity in CH
Cygni is driven by accretion through a disk with a thermal-viscous
instability, similar to the instabilities believed to exist in dwarf
novae and suggested for FU Ori pre-main-sequence stars and soft X-ray
transients.
\end{abstract}




\section{Introduction}

Observationally, the symbiotic star CH Cygni is a very complex system
\citep[see][for a review of its properties]{ken01}.  It contains an
accreting white dwarf fed from the wind of a red giant and surrounded
by an ionized nebula.  There is debate over whether CH Cygni is a
double or a triple system.  It displays at least 3 or 4 long-period
photometric or radial velocity variations ($\sim$100 d for the red
giant pulsation; $\sim$760 d from either an inner binary orbit or
repetition of short $U$-band activity drop-outs and radial velocity
variations; 5200-5700 d for the binary orbit, or outer stellar orbit;
possible 32-yr $JHK$ variation).  In addition to collimated jets, the
hot component also produces a less-collimated outflow during periods
of activity
\citep{eyres02,sko02inprep}.  Episodes of dust condensation further
complicate analysis of the stellar components.

\begin{figure*}
\begin{center}
\epsfig{file=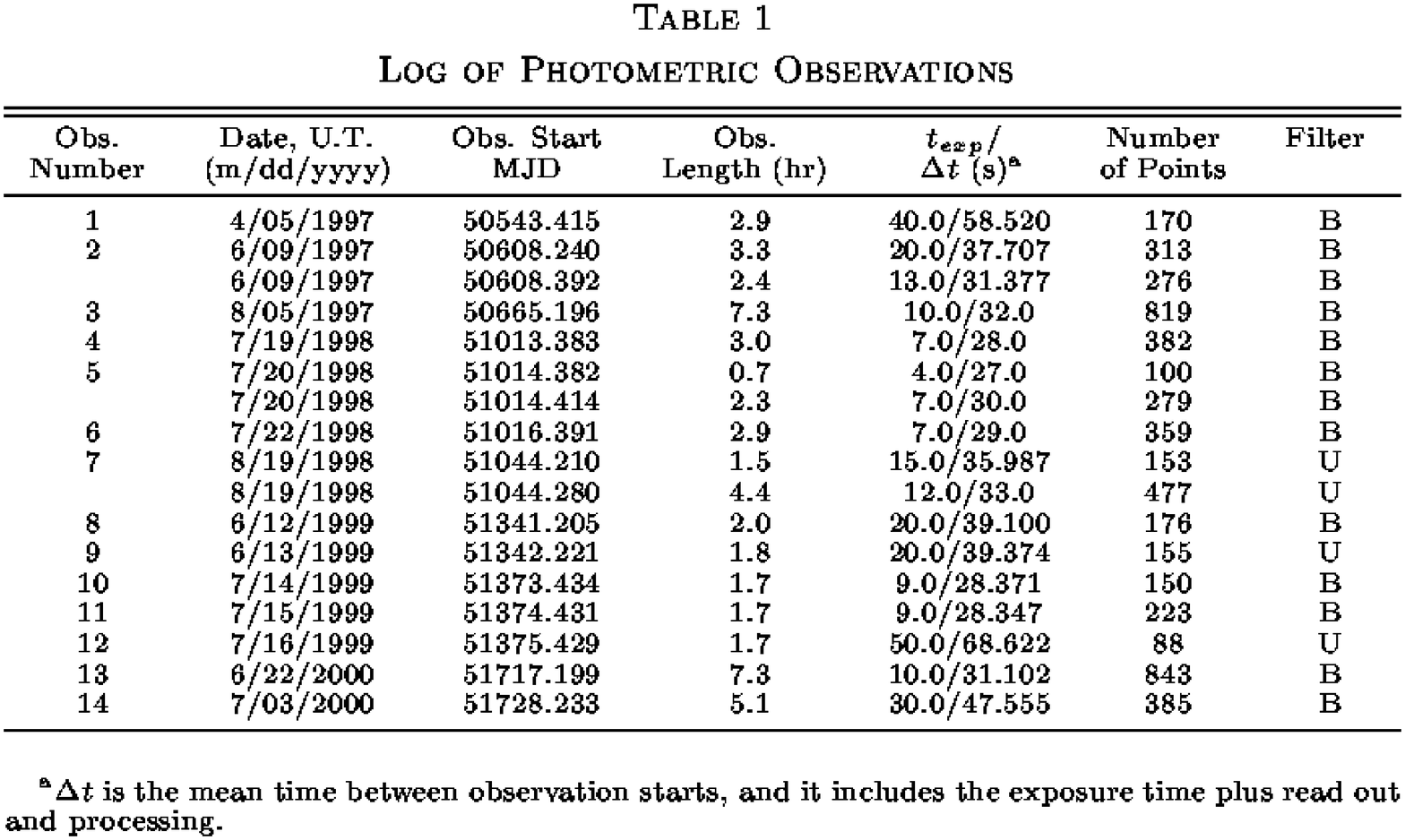,width=14cm}
\end{center}
\vspace{-1cm}
\end{figure*}

In the face of this complexity, one way to isolate emission from the
accretion region 
is to examine rapid (time scales of minutes to hours) optical
variations.
It is very unlikely that the red giant in CH Cygni could produce
variations on a time scale of minutes or seconds, given a dynamical time
$t_{dyn} \sim (R^3/GM)^{1/2}$, which is generally greater than 0.5 d.
In fact, high-time-resolution
$U$-band photometry of CH Cygni by many authors \citep[e.g.,][and
references therein]{mski90} indicates that that the rapid variations
come from the hot component,
and not the red giant.  In another symbiotic star, T CrB, \cite{zb98}
found flickering ``indistinguishable'' from that seen in cataclysmic
variables (CVs), and they
concluded that it is therefore probably also produced in the vicinity
of the WD.  

In this paper, we investigate the role of the accretion disk in 
optical brightness-state changes in CH Cygni.  We performed rapid
optical photometry and optical spectroscopy for CH Cygni between 1997
and 2000.
Our observations cover several interesting phases.
In 1997, CH Cygni produced a radio jet after a 
drop in the optical flux \citep[][observations from this period are
discussed in Sokoloski \& Kenyon 2003; hereafter paper I]{kar98}.  In
1998, it entered a high activity state; in 1999, the hot source was
eclipsed by the red giant
\citep[according to][]{eyres02}.  The
properties of the optical flickering changed during each of these
phases, and in combination with the optical spectra, these changes
provide information about the accretion disk during jet production,
the nature of the high activity state, and the timing of the eclipse.

This paper is divided into six sections.  After the initial
description of the observations in \S\ref{sec:obs}, we report the
results from our fast photometry in \S\ref{sec:results}. In
\S\ref{sec:eclipse}, we discuss the disappearance of all minute- to
hour-time-scale variations during the proposed eclipse of the
accreting WD by the red giant in 1999.
In \S\ref{sec:mag}, we describe the timing analysis of the light
curves.  No coherent or quasi-periodic oscillations were found, and we
discuss
the difficulty for magnetic propeller models of CH Cygni in the face of
this lack of evidence for a strong magnetic field.
In \S\ref{sec:activity}, we suggest that the activity in CH Cyg may
instead be driven by an unstable accretion disk.

\section{Observations and Data Analysis} \label{sec:obs}

We performed high-time-resolution optical differential photometry at
$B$ and at $U$, using the 1-meter Nickel telescope at Lick
Observatory, on 14 nights between 1997 and 2000
(see Sokoloski, Bildsten, \& Ho 2001 for description of observing
technique and instruments).
Eclipses in the best comparison star, SAO 31628 (see Sokoloski
\& Stone 2000), affected 3 of the 14 observations.
To obtain a uniform, high-quality set of light curves with this
comparison star, we divided its light curves on 1998 July 22 and 2000
June 22 by a 6th-order-polynomial fit to the eclipse profile.  This
correction added only a small amount of power to the power spectrum of
the CH Cygni observations, as measured by computing the power spectrum
of the ratio of SAO 31628 and a second comparison star in the field.
This slight extra power only appeared at the lowest frequencies, and
did not hamper our search for coherent oscillations.  It also did not
diminish our ability to measure the power-law slope on 22 July 1998
since it was insignificant compared to the actual broad-band power due
to intrinsic variations from CH Cygni.  On 1999 July 3, the SAO 31628
eclipse only affected the last 80 minutes of a 6-hr light curve, so
the affected portion of the data was not used.  Table~1 lists the
photometric observations.

\begin{figure*}
\begin{minipage}{18cm}
\plottwo{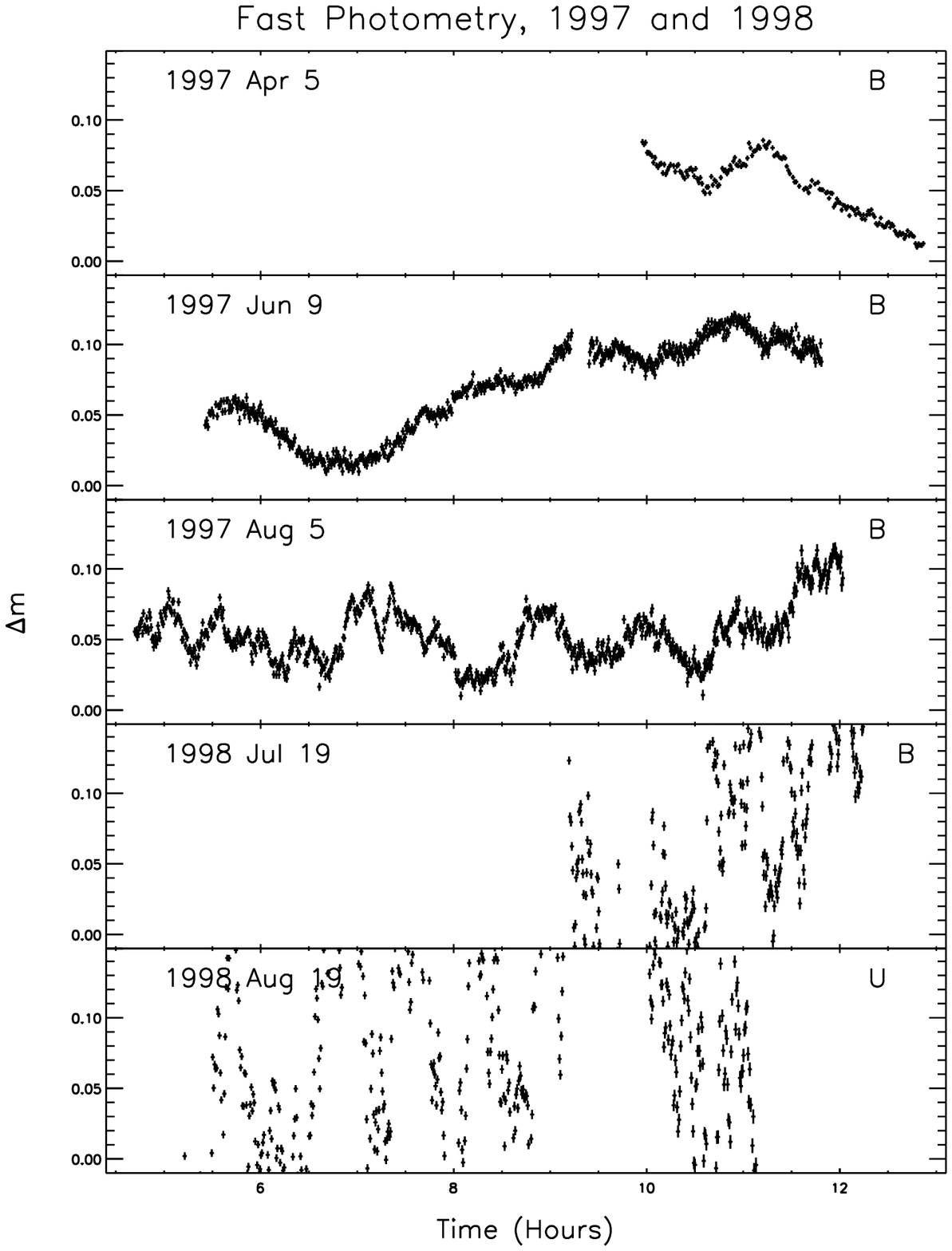}{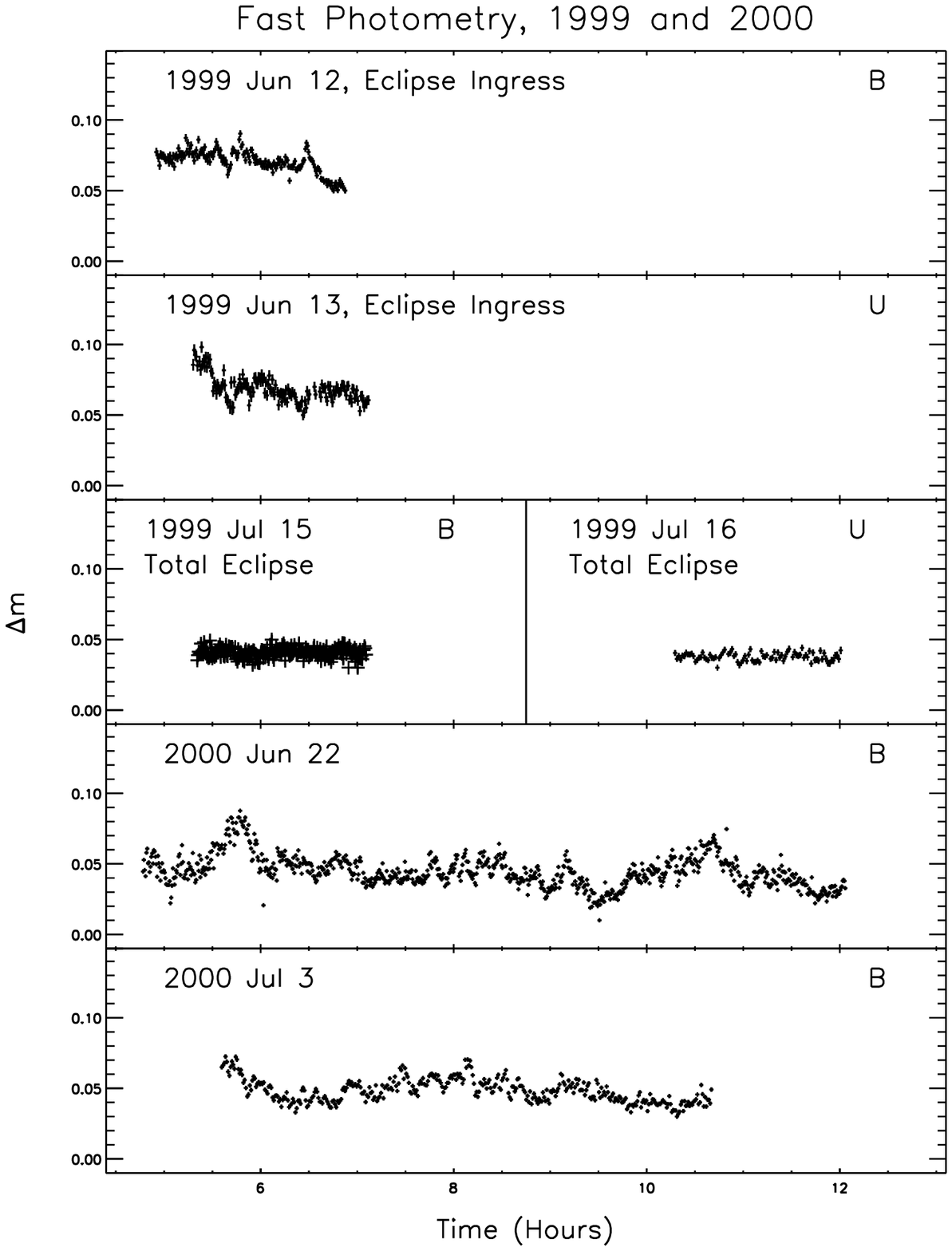} 
\caption{\footnotesize Example CH Cyg light curves from 1997 to
2000.  All data are plotted on the same scale for direct comparison
(the amplitude of the variations in 1998 span three times the
magnitude range shown; full high-state light curves are shown in
Fig. \ref{fig:lcsb}).  Low-amplitude flickering is present in 1997 and
2000.  All variations disappear during the eclipse of the hot
component in 1999.
\label{fig:alllcs}}
\end{minipage}
\end{figure*}

P. Berlind, M. Calkins, and several other observers acquired low
resolution optical spectra of CH Cygni with FAST, a high throughput,
slit spectrograph mounted at the Fred L. Whipple Observatory 1.5-m
telescope on Mount Hopkins, Arizona \citep{fab98}.  They used a 300 g
mm$^{-1}$ grating blazed at 4750 \AA, a 3\arcsec~slit, and a thinned
Loral 512 $\times$ 2688 CCD.  These spectra cover 3800--7500 \AA~at a
resolution of 6 \AA.  We wavelength-calibrated the spectra in NOAO
IRAF\footnote{IRAF is distributed by the National Optical Astronomy
Observatory, which is operated by the Association of Universities for
Research in Astronomy, Inc. under contract to the National Science
Foundation.}.  After trimming the CCD frames at each end of the slit,
we corrected for the bias level, flat-fielded each frame, applied an
illumination correction, and derived a full wavelength solution from
calibration lamps acquired immediately after each exposure.  The
wavelength solution for each frame has a probable error of $\pm$0.5
\AA~or better.  To construct final 1-D spectra, we extracted object
and sky spectra using the optimal extraction algorithm APEXTRACT
within IRAF.  Most of the resulting spectra have moderate
signal-to-noise, S/N $\gtrsim$ 30 per pixel.

\section{Photometric Results} \label{sec:results}

Our high-time-resolution light curves 
demonstrate tremendous diversity in both amplitude and frequency
content (see Figure~\ref{fig:alllcs}).
The observations from 1997 show low-amplitude ($\Delta m < 0.15$ mag)
variations that initially are unusually smooth. 
The evolution of the light curve in 1997, when a radio jet was
produced, is discussed in paper I.
In 1998 July and August, we see full, large-amplitude
($\Delta m \sim 0.5$ mag) CV-like flickering when CH Cygni is in a
high state.  Figure~\ref{fig:lcsb} shows the four high-state light
curves from 1998.
In 1999 July, the optical flux is constant to the level of a few
tenths of a percent.  Finally, in 2000 June and July, we again see
low-amplitude flickering after CH Cyg returns to an optical low state.
The optical brightness levels we refer to as ``high,'' ``low,'' and
``very low'' are marked on the long-term AAVSO light curve shown in
Figure \ref{fig:ltlc}.  Figure~\ref{fig:pdss} shows example power
spectra from the low-state, high-state, and eclipse light curves
(excluding the unusual light curves from early 1997, which are
discussed in paper I).  The power spectra were generally well-fit with
power-law plus a constant (for the high-frequency white noise) models,
$P = Af^{-\alpha} + B$, with power-law index $\alpha = 1.8 \pm 0.1$.
In the 2000 low state, the power-law was slightly flatter, with
$\alpha \approx 1.6$ on 2000 June 22, and $\alpha \approx 1.3$ on 2000 
July 3.

Table 2 lists two measures of the overall variability
amplitude for each observation (except observation \#10, on 1999 July
14, for which the weather was poor).  The fractional root-mean-square
(rms) variation (in mmag) is in column 2.  Column 3  lists the
fractional variation multiplied by the approximate $B$-band luminosity
of CH Cygni at the relevant epoch (taking a bandwidth of 1000 \AA) to
provide an estimate of the optical luminosity of the variable
component, $L_{B,var}$.

\section{Disappearance of Flickering During the Long-Period
Eclipse} \label{sec:eclipse}

There has been much debate over whether CH Cygni is a double or a
triple stellar system (e.g., in favor of triple: Hinkle 1993; Skopal
1995; Skopal et al. 1996; Iijima 1998; in favor of double:
e.g., Miko{\l}ajewska 1994; Munari et al. 1996; Ezuka, Ishida,
\& Makino 1998).  The most controversial contention is that the 756-d
radial velocity variations are orbital, in addition to those from the
more generally accepted 14-yr orbital period.
Our conclusions in this paper
do not depend upon whether CH Cygni is a double or triple system.

\begin{figure*}
\begin{center}
\epsfig{file=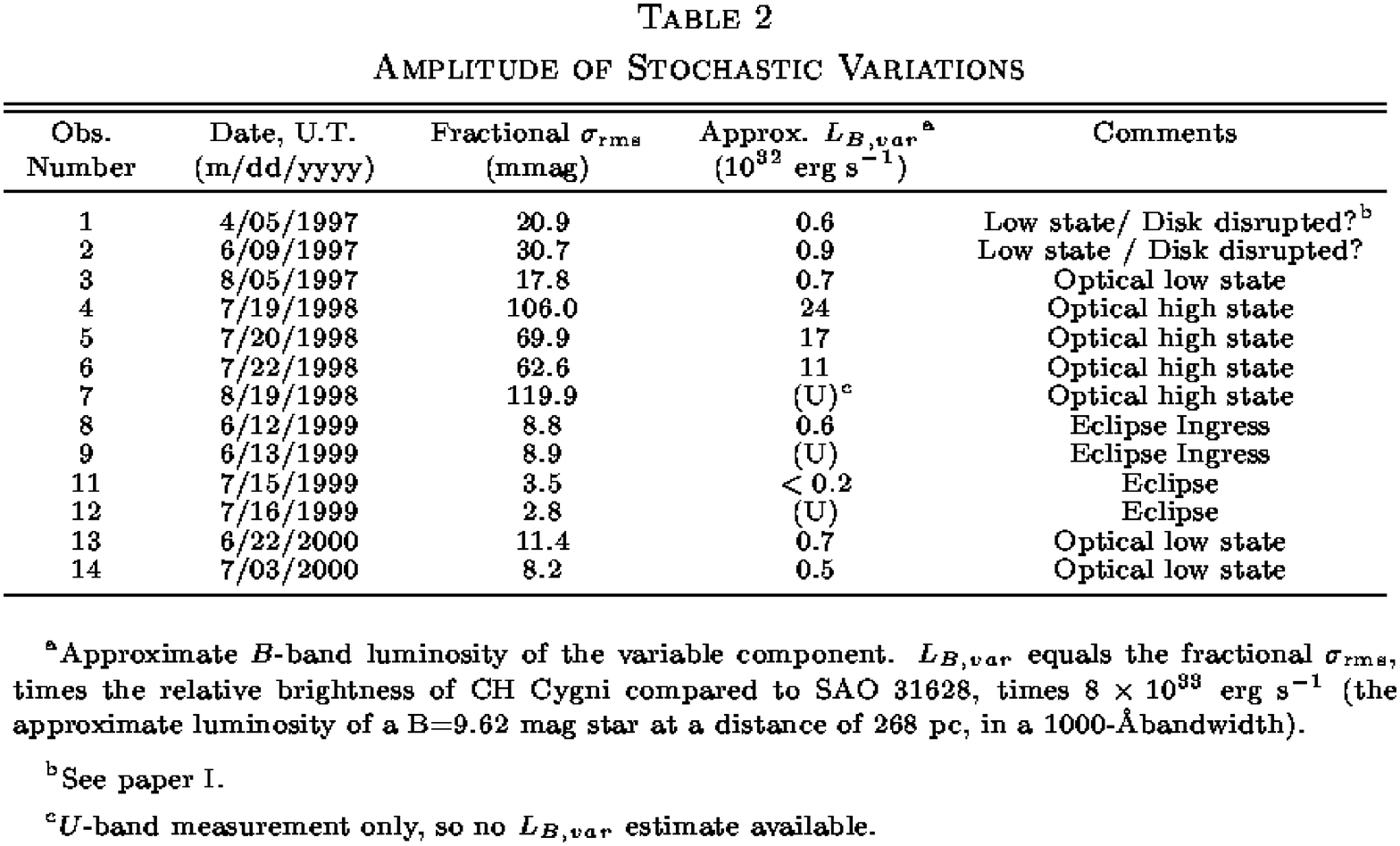,width=14cm}
\end{center}
\vspace{-0.8cm}
\end{figure*}

The light curves from our 1999 observations show very little rapid
variability.  In 1999 June, we see low amplitude flickering
($\sigma_{rms} < 9$ mmag).  The 1999 July light curves are constant at
our sensitivity limit of a few mmag.  Our measurement of a very tight
flickering upper limit in 1999 July is consistent with both the
interpretation by \cite{eyres02} of this event as an eclipse by a
companion orbiting with a period of roughly 14 years (and their timing
of the eclipse as beginning in early June), and our assumption that
the flickering is due to accretion on to the WD.  Their evidence for
the eclipse includes a deep $U$-band minimum, disappearance of the
broad bases of the Balmer emission lines, and the close coincidence of
this behavior with spectroscopic conjunction.  The eclipse began in
early 1999 June, and the minimum of the eclipse occurred at JD
$2451426 \pm 3$
\citep{eyres02}. 
We detected some variations in 1999 June, when the optical light curve
had not quite reached eclipse minimum; the beginning of totality must
have fallen between mid-June and mid-July.  According to
\cite{eyres02}, the $U$-band flux dropped in about 10 days, beginning
in early 1999 June.  Some portion of the disk may therefore still have
been visible on 1999 June 12 and 13.

In Fig.~\ref{fig:eclipsespec}, a comparison of spectra from 
before and during the proposed eclipse confirms that the hot component
disappeared.  During the eclipse, the blue veiling continuum vanished,
and the red-giant absorption features dominated the spectrum.  The
He\,I $\lambda$5015\AA\, emission line from the region close to the WD
also disappeared, while the lower-excitation H$\alpha$, H$\beta$, and
[O\,III] $\lambda$5007\AA\, emission lines weakened.  The lowest
excitation [O\,I] $\lambda$6300 emission line, which probably comes
from the outer-most, low density region of the nebula, remained
unchanged.  The overall line changes are consistent with an eclipse of
the accreting WD and inner nebula by the red giant.

\section{The Question of Magnetism} \label{sec:mag}

Coherent oscillations in optical or X-ray emission are one of the most
decisive signatures of magnetic accretion in CVs and X-ray pulsars.
Magnetically channeled accretion produces hot spots 
on the surface of the compact star that rotate in and out of the line
of sight at the WD or neutron-star spin period.  In CVs, X-rays from
the rotating accretion column can act like a lighthouse beam
illuminating the disk and/or stellar surface.
The optical pulsation amplitudes due to magnetic accretion in
intermediate polars generally range from a few percent to 10 - 20 \%
(Warner 1995). The optical oscillation amplitude for the well-known
magnetic propeller AE Aqr is slightly smaller: 0.2 - 0.3\% when it is
quiet, and about 1\% during flares
\citep{pat79}.  Its X-ray emission, however, is modulated with an
amplitude of 25\% \citep{war95}. 

There is currently only one symbiotic binary with convincing evidence
for a strongly magnetized WD.  Sokoloski \&
Bildsten (1999) discovered the first
symbiotic magnetic accretor in the prototypical symbiotic Z
Andromedae.  They repeatedly detected a statistically significant
oscillation at $P = 1682.6 \pm 0.6$ s, which they interpreted as the
spin period of the accreting white dwarf.  Thus, coherent brightness
oscillations are also a signature of magnetic accretion in symbiotic
stars.

\begin{figure*}
\begin{center}
\epsfig{file=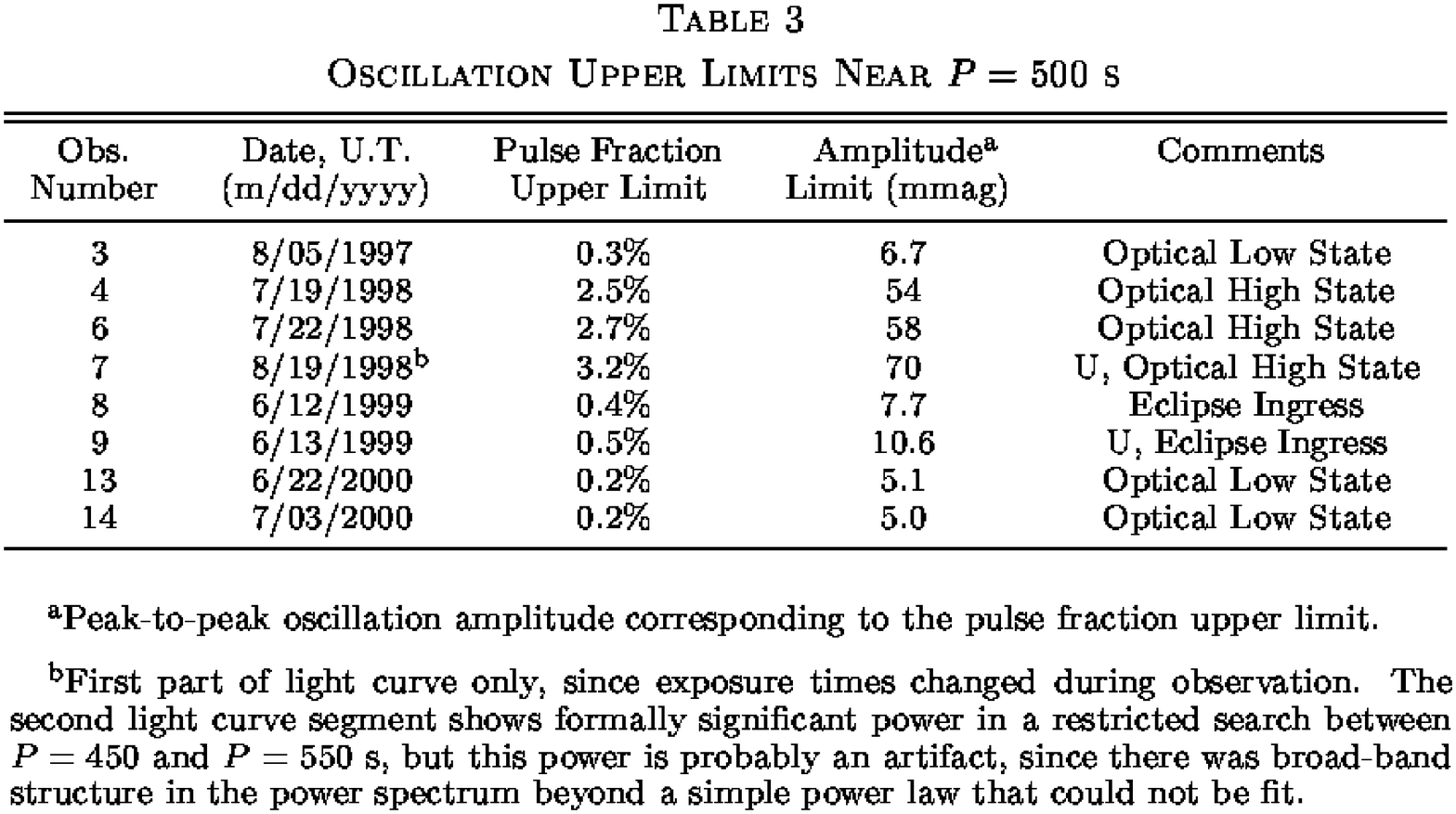,width=14cm}
\end{center}
\vspace{-0.8cm}
\end{figure*}

\begin{deluxetable}{ccccc}
\tabletypesize{\footnotesize}
\tablecaption{Oscillation Upper Limits Near $P = 500$ s \label{tbl:osclims}}
\tablewidth{0pt}
\tablehead{
\colhead{Obs.} & \colhead{Date, U.T.} & \colhead{Pulse Fraction} &
\colhead{Amplitude\tablenotemark{a}} & \colhead{Comments} \\
\colhead{Number} & \colhead{(m/dd/yyyy)} & \colhead{Upper Limit} &
\colhead{Limit (mmag)} & \colhead{} } 
\startdata
3 & 8/05/1997 & 0.3\% & 6.7 & Optical Low State\\
4 & 7/19/1998 & 2.5\% & 54 & Optical High State\\
6 & 7/22/1998 & 2.7\% & 58 & Optical  High State\\
7 & 8/19/1998\tablenotemark{b} & 3.2\% & 70 & U, Optical  High State\\
8 & 6/12/1999 & 0.4\% & 7.7 & Eclipse Ingress \\
9 & 6/13/1999 & 0.5\% & 10.6 & U, Eclipse Ingress\\
13 & 6/22/2000 & 0.2\% & 5.1 & Optical Low State\\
14 & 7/03/2000 & 0.2\% & 5.0 & Optical Low State\\
\enddata
\tablenotetext{a}{Peak-to-peak oscillation amplitude corresponding to the pulse
fraction upper limit.}
\tablenotetext{b}{First part of light curve only, since exposure
times changed during observation.  The second light curve segment
shows formally significant power in a restricted search between
$P=450$ and $P=550$ s, but this power is probably an artifact, since
there was broad-band structure in the power spectrum beyond a simple
power law that could not be fit.}
\end{deluxetable}

\begin{inlinefigure}
\vspace{0.5cm}
\plotone{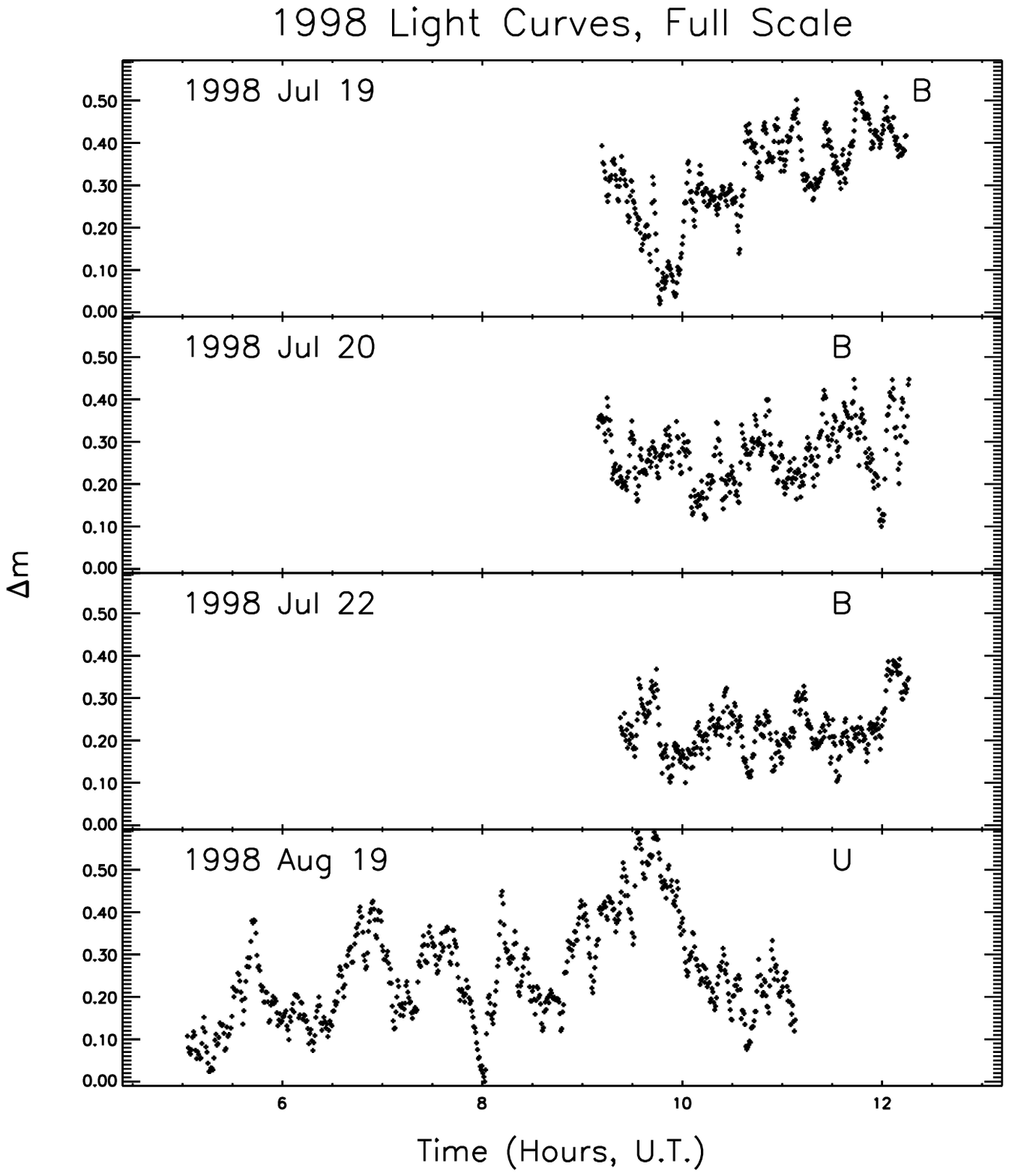}
\caption{\footnotesize Complete set of light curves from the large-amplitude
flickering period in 1998.  The ordinate for each of these plots spans
0.6 mag (compared to 0.15 mag in Figure 1).
\label{fig:lcsb}}
\end{inlinefigure}

To explain the jets and different brightness states in CH Cygni,
\cite{mm88} proposed a magnetic propeller model, based on the oblique
rotator theory of Lipunov (1987)\nocite{lip87}.  They suggested that
the inner disk is ejected and a jet produced when the accretion rate
onto the WD drops and the system changes from the accretor to the
propeller state.
In their model, the optical flickering is due to the interaction of
the accreted material with a strong magnetic field.  Later,
\cite{mski90} reported the detection of an oscillation with a period
of 500 s, which they claimed was the rotation period of the white
dwarf.  If confirmed, this oscillation would provide some support for
the magnetic propeller model for CH Cygni
\citep[which is discussed further by e.g.,][]{mmk90,mski90,panmski00}.

\begin{inlinefigure}
\vspace{0.5cm}
\plotone{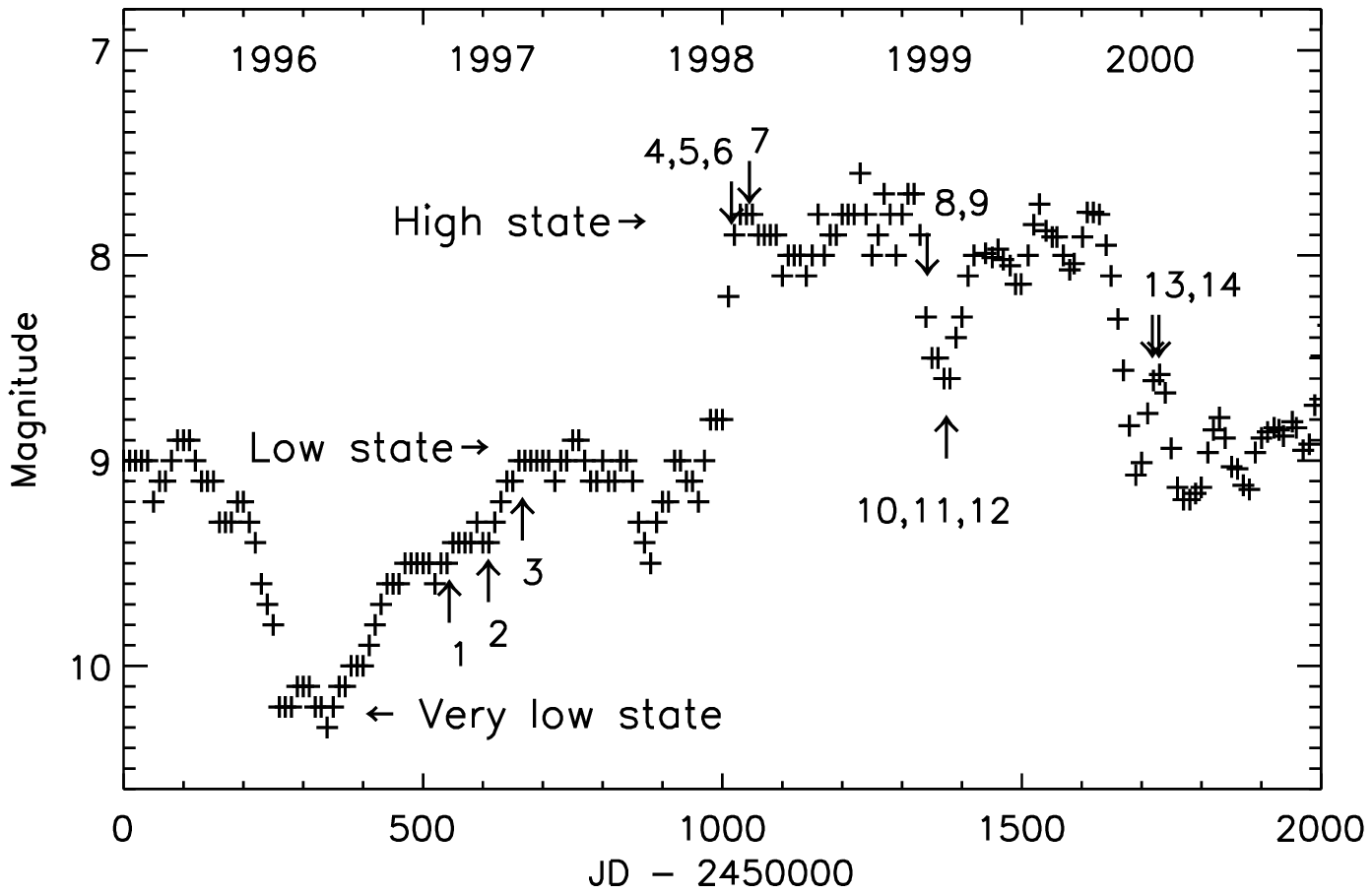}
\caption{\footnotesize Long-term optical light curve of CH Cygni,
from the AAVSO.  The times of our 14 flickering observations are
marked with arrows.
\label{fig:ltlc}}
\end{inlinefigure}

X-ray, radio, and optical observations by other authors, however,
do not support the magnetic interpretation of CH Cygni. \cite{ezu98}
observed stochastic X-ray variations on time scales as short as 100 s,
but did not detect any coherent oscillations.  Also, their fit to the
X-ray spectrum
yielded $kT = 7.3 \pm 0.5$ keV, which they claim is low compared to
the $kT = 10 - 40$ keV typically found for magnetized CVs
\citep{if95}.  By assuming equipartition of energy in the 1986
radio jet,  
\cite{crock01} estimated 
a magnetic field strength in the jet
that is consistent with a WD
surface field strength of only 10 G.  In addition, \cite{rod97} and
\cite{hoard93} found no evidence of a period between 500 and 600 s in
their optical photometry for CH Cygni (although they did 
\begin{inlinefigure}
\plotone{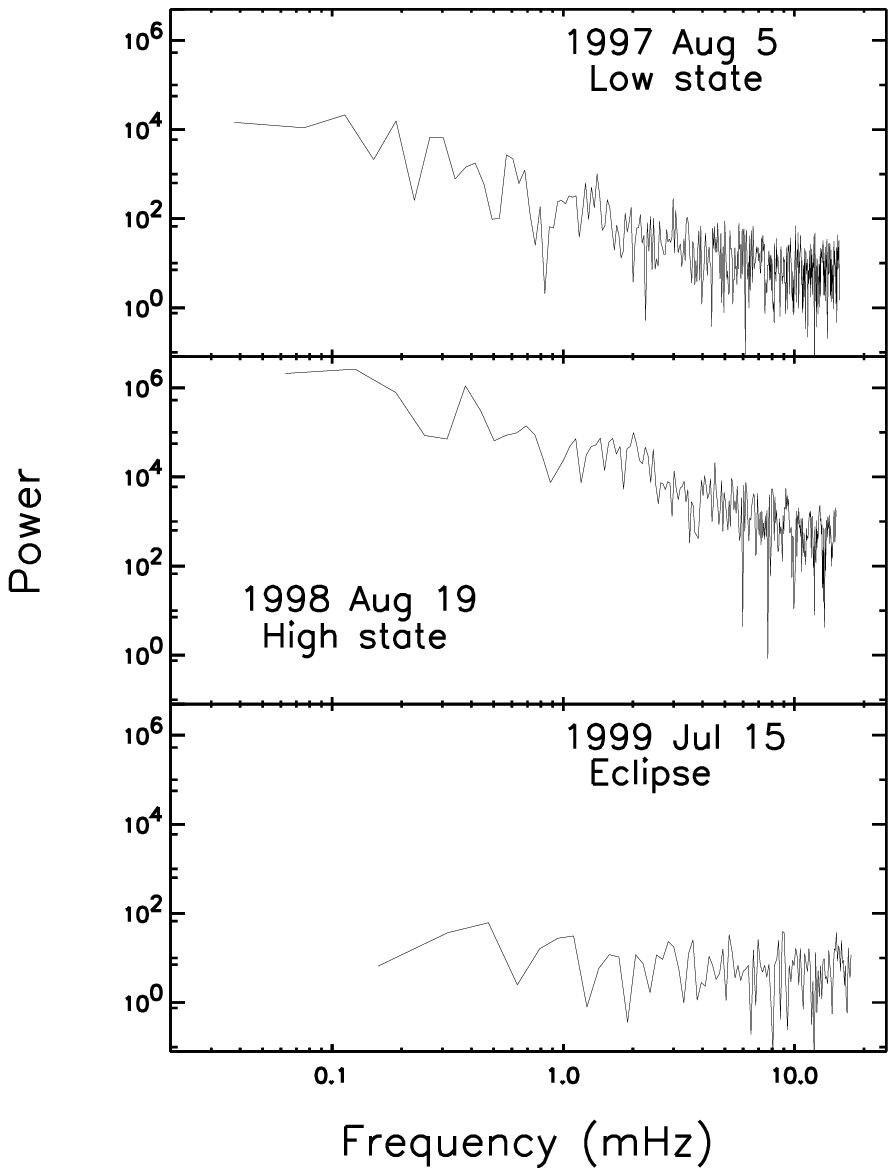}
\caption{\footnotesize Example power spectra (normalized to the total
number of source counts in each observation) from 
1997, 1998, and 1999.
The power is greater at all frequencies in 1998 August, when the
system was in an active state.  The bottom panel shows the lack of
variations during WD eclipse. \label{fig:pdss}}
\end{inlinefigure}

\noindent
claim to find
periods at approximately 2200 and 3000 s).  

To search for oscillations that would indicate magnetic accretion in
CH Cygni, we computed power spectra for each of the light curves in
our data set.  We searched for statistically significant excess power
above the broad-band red-noise power, and found no coherent or
quasiperiodic oscillations (see Appendix~\ref{app:timing} for a
discussion of oscillation detection statistics when a source also
produces stochastic variations).  Even in 1998, when CH Cygni was in a
high state - presumably due to accretion onto the WD - and an
oscillation would have been most likely to be present in the magnetic
model, we did not find any oscillations.

By performing Monte Carlo simulations, we placed upper limits on the
amplitude of any sinusoidal variations in the data.  To make this
test, we
added oscillatory
signals to our real data and measured the amplitude needed to
produce a detectable signal.  Because we searched for
oscillations against a background of broad-band (``red'') power, which
is greater at lower frequencies, the upper limits
are also a function of frequency.  For example, 
in our 19 July 1998 observation, we could detect an oscillation with a
semi-amplitude of 2\% or larger if its period was between 60 and 300
s.  If the period was between 2500 and 3500 s, we could only constrain
the semi-amplitude of any oscillation hidden in the flickering to be
less than 17\%.

Table~3 lists the results of searching the period
range of 450 to 550 s for all the light curves where the power
spectrum could be well characterized by a power-law model\footnote{The
oscillation detection statistics are not meaningful unless there is an
acceptable fit for the functional form of the broad-band power.  See
Appendix A.}.  We list oscillation amplitudes that were detectable at
a 95\% confidence level in 95\% of the realizations.  They range from
0.2\% in the low states in 1997 and 2000 to 3.2\% in the 1998 high
state.
To reduce the variance of the noise powers at high frequencies, and
therefore improve our sensitivity to a possible quasi-periodic
oscillation (QPO) near 500 s, we divided each light curve into 2 - 10
segments, and averaged the power spectra from these individual
segments.  Applying this procedure, we did not find statistically
significant QPOs above the background ``red'' power spectrum for any
of the observations of CH Cygni listed in 
Table~3.

\begin{inlinefigure} 
\plotone{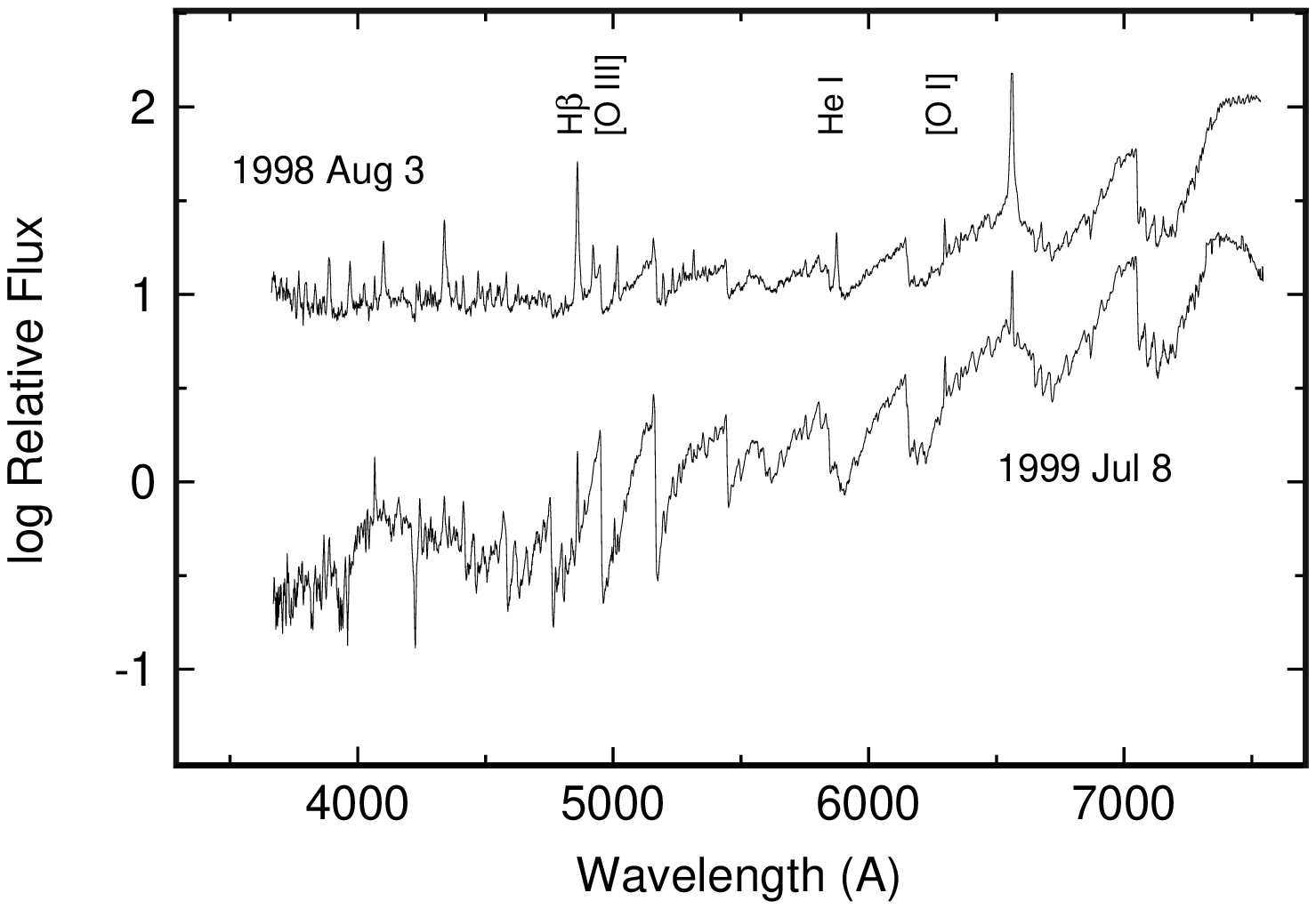}
\caption{\footnotesize Top: high state spectrum, before 
the proposed eclipse.  Bottom: spectrum during the eclipse.  In the
eclipse spectrum, the hot component is absent and the high-excitation
He\,I emission line is gone, whereas the lower excitation lines from
the outer nebula remain. \label{fig:eclipsespec}}
\end{inlinefigure}

The non-detection of coherent oscillations at any frequency tested in
either the high or low state, our limits on coherent oscillations near
500 s, and the lack of evidence for any QPOs near this period make the
magnetic interpretation for CH Cyg questionable.
As for the theoretical necessity of a strong magnetic field for jet
production, \cite{fh00} found, for X-ray binaries, that the truncation
of the inner disk by a strong magnetic field could actually inhibit
the formation of a jet, not cause it.
Therefore, we conclude that alternatives to the magnetic propeller
model for CH Cygni should be explored.

\section{Implications for Brightness-State Changes} \label{sec:activity}

In almost every type of interacting binary star (or other accreting
system), disk accretion produces stochastic brightness variations.
The flickering in CH Cygni disappeared during eclipse of the hot
component (see \S\ref{sec:eclipse}), and must therefore come from a
region physically near the WD.  Inhomogeneities in and around an
accretion column in polars can produce stochastic variations
\citep[e.g.,][and references therein]{war95}, and
\cite{mm88} have suggested that brightness fluctuations in CH Cygni are due to
interaction of the accreted material with a strong magnetic field.
But no spin modulation has been found in CH Cygni (see
\S\ref{sec:mag}), so the flickering in this system is unlikely to be
related to magnetism.  
\cite{lt87} noted that the X-ray emission detected by EXOSAT
could be due to either a disk boundary layer or shock-heated colliding
winds.  Fast stochastic variations \citep[][later confirmed variations
with time scales as short as 100 s]{ezu98} support a disk boundary
layer.  Theoretically, a disk can form from the red-giant wind in CH
Cygni if the wind speed is $\la 50$ km s$^{-1}$ (for $P_{orb} \approx
760$ d) or $\la 30$ km s$^{-1}$ \citep[for $P_{orb} \approx 14$ yr;
taking a total system mass of $2 \msun$ in each case;][]{liv88}.
\cite{dyk98} measure a stellar-disk size for the red giant in CH Cygni
of 10.4 mas.  For a distance of 245 $\pm 50$ pc (from the mode of the
Hipparcos parallax probability distribution), this angular size
corresponds to a stellar radius of approximately 270 $\rsun$, which is
reasonable for an M6-7III giant. \cite{schild99} also estimate a
radius of $280 \pm 65\, \rsun$, from the $J$ and $K$ magnitudes.  The
escape speed is therefore about 40 km s$^{-1}$.  The red-giant wind
speed is probably less than the escape speed, and so it is reasonable
to expect a disk to form.  Thus, the most natural explanation for the
fast fluctuations is disk flickering, and changes in the flickering
can tell us about changes in the disk.

Looking back at Table~2, we see that the optical luminosity of the
rapidly variable component, $L_{B,var}$, increased by more than a
factor of 20 between 1997 and the optical high state in 1998.  Given
this large increase, the physical process that produces the optical
flickering must have been largely responsible for the overall optical
brightening of CH Cygni (by roughly a factor of 7) between 1997 and
1998.  The large increase in the luminosity of the flickering
component therefore implies that the change to a high state was due to
an increase in the accretion rate through the disk, $\mdot$.

One way to produce a change in $\mdot$ through a disk is via a
dwarf-nova-like thermal-viscous disk instablity
\citep[e.g.,][and references
therein]{war95}. The disk in CH Cygni
could be much larger than disks in CVs, 
the time-averaged accretion rate could be higher,
and it is formed from a wind rather than a stream. Duschl (1986a,b)
examined large symbiotic accretion disks theoretically, and found that
they are also likely to experience limit-cycle instabilities.
The $\mdot$-$\Sigma$ relations (where $\Sigma$ is the surface density)
have negative-sloping sections
in the temperature region where H is ionized, as in CVs, and also at
lower temperatures, where the molecular opacity
changes\footnote{Duschl (1986a,b) considered large disks in
main-sequence symbiotic stars, but his results are also relevant for
WD symbiotics.
}. 
Again taking a distance of 245 $\pm 50$ parsec for CH Cygni, $M_V \sim
-1$ in the 1979-1984 high state, $M_V
\sim 1$ in the low state after the 1985 jet, and $M_V
\sim 4$ at the extremely low point after the jet in 1997. 
These values are reasonable for dwarf-nova high and low states given
that the disk in CH Cygni is likely to be large and that the red giant
makes a significant contribution at $V$.  $L_{B,var}$ is of the order
of the luminosity expected from accretion onto a WD ($\sim 10^{32} -
10^{33}$ erg s$^{-1}$), with a lower accretion rate onto the WD in
1997 compared to 1998.  Furthermore, the expected recurrence time for
an instability in a large disk, assuming an average accretion rate of
$10^{-8}\msyr$ and an accretion disk radius on the order of tens of
solar radii, is on the order of years \footnote{The recurrence time is
either approximately the viscous time at the outer edge of the disk
for an outside-in outburst, or given by expression (3.29a) in Warner
1995, from \cite{csw88} for an inside-out outburst.}.  This expected
recurrence time agrees well with the time scale of state changes in CH
Cygni.  Finally, in the high state, the hot component in CH Cygni has
an F-type supergiant spectrum, very similar to the high-state disks of
dwarf novae in outburst.

There are, however, some major differences between the behavior of CH
Cygni and that of dwarf novae in the optical.  Instead of having the
fast rise and exponential decay shape common to dwarf-nova outbursts, the high
states in CH Cyg tend to be plateau-like.  CH Cygni's long-term light
curve is also much more complex than long periods of quiescence with
outbursts superimposed.  

\cite{bogtar01} explored an alternative hypothesis - that the
long-term optical variations in CH Cygni (time scale of months to
decades) are due to 
obscuration by dust.
Several observations after the 1984 and 1996 optical flux declines
revealed an increase in the column density of dust
\citep[e.g.,][]{ty88,ty92,mun96,bogtar01}. 
The 1996 optical fading to the very-low state, however,
occurred 100 to 150 days before the increased production of dust
\citep[as indicated by a sudden change in $J-H$ color;][]{bogtar01},
and the 1984 optical fading to the low state also occurred before the
dust condensation in 1985-1987 \citep{ty88}.  Furthermore, the
spectral changes described in paper I are consistent with a decrease
in the ionization state of the nebula, not obscuration of the entire
system.  
Therefore, 
the optical fadings (and associated jet events) were either unrelated
to the episodes of dust condensation, or even somehow caused the dust
production.

The increase in the optical flickering luminosity in the high state
indicates that the activity in this system is accretion-driven.  Since
the disk in CH Cyg (assuming one exists) is expected to be subject to the
same instability as in dwarf novae \citep[and possibly
another;][]{duschl86a,duschl86b}, we propose that such instabilities
drive the activity in CH Cygni as well.  Since jets are sometimes
associated with state changes in CH Cygni, collimated outflows could
be related to the disk instabilities.  An alternative model, in which
material is expelled by a rapidly rotating magnetic field, requires
that the WD have a strong magnetic field.  We do not find evidence for
such a field, and therefore favor the disk-instability scenario.

\acknowledgments

This work has been supported in part by NSF grant INT-9902665.  In
this research, we have used and acknowledge with thanks, data from the
AAVSO International Database, based on observations submitted to the
AAVSO by variable star observers worldwide.  We also thank
R. P. S. Stone for performing the photometric observations in 2000,
and C. Brocksopp, M. Muno, and L. Bildsten for helpful comments.


\appendix
 
\section{Oscillation Detection in Light Curve with Broad-Band Power}
\label{app:timing} 

The presence of large-amplitude stochastic variations
makes coherent oscillations, which are the hallmark of magnetic
accretion, difficult to identify.
In order to make any statistical statements about coherent
oscillations, QPOs, or the quality of a model fit to the power
spectrum, one must assume some distribution of the noise powers about
the ``true'' power spectrum shape .  Van der Klis (1989) found that
for the neutron-star binary GX 5-1, which flickers and has a QPO in
the X-ray regime, the distribution of noise powers at a given
frequency is well described by a $\chi^2$ distribution about the mean
value of the power at that frequency (averaged over thousands of
observations).
\citet{db82} predicted a $\chi^2$ noise-power distribution on
theoretical grounds for a ``red'' spectrum that is the integral of
white noise.  Since we do not have thousands of observations, we take
the suggestion of \citet{vdk89} and assume a $\chi^2$ distribution
about the best fit power law for each observation.
The probability $\Pr (P_{noise} > P)$ that the noise power in a single
frequency bin will exceed the threshold value $P$ is 
\begin{equation}
\Pr (P_{noise} > P) = Q(\chi^2 | \nu),
\end{equation}
with $\chi^2 = 2P/P_{fit}$ and $\nu = 2$, 
where the $\chi^2$ distribution with $\nu$ degrees of freedom is given
by
\begin{equation}
Q(\chi^2 | \nu) = \left[ 2^{\nu/2}\, \Gamma(\nu/2) \right]^{-1}
\int_{\chi^2}^{\infty} t^{(\nu/2) - 1}\, e^{-t/2}\, dt,
\end{equation}
and $P_{fit}$ is the value of the model fit at the frequency of
interest.  Performing the integral, we get the well-known result $\Pr
(P_{noise} > P) = e^{-P/P_{fit}}$.  For a search of multiple frequency
bins, the probability $\Pr (P_{noise} > P)$ that at least one of the
noise powers will exceed $P$ is
\begin{equation} \label{eqn:mfreqsearch}
\Pr (P_{noise} > P) = 1 - \left(1 -e^{-P/P_{fit}} \right)^{n_{freq}},
\end{equation}
where $n_{freq}$ is the number of frequencies searched.  A peak with
power $P$ in the power spectrum only indicates the presence of a
coherent oscillation if this probability is small.

The original, full frequency-resolution power spectra provide the best
sensitivity to a narrow (single frequency bin) feature in the power
spectrum.  Sensitivity to a broad feature (QPO), on the other hand,
improves when power spectra produced from light curve segments are
averaged, as long as the width of the feature is larger than one
frequency bin in the averaged power spectrum (van der Klis 1989).
Using the notation of \cite{vdk89}, averaging $M$ power spectra
reduces the variance in the combined power spectrum by a factor of
$MW$, where $W$ is an additional factor for binning in the frequency
domain.  The noise powers in the averaged and binned power spectrum
are distributed like $\chi^2$ with $2MW$ instead of 2 degrees of
freedom, and $\chi^2=2MWP/P_{fit}$ \citep[see][equation 3.4]{vdk89}.
Doing the $\chi^2$-distribution integral in the general case, we find
the probability for a noise power to exceed a threshold value $P$ in a
single frequency bin of a binned and averaged power spectrum:
\begin{equation}
(P_{noise} > P) = Q(2MWP/P_{fit}| 2MW) = e^{-MWP/P_{fit}} \; \sum_{n=1}^{MW}
\frac{(MW)^{MW-n}(P/P_{fit})^{MW-n}}{(MW-n)!}. 
\end{equation}
The expression for a
multiple-frequency-bin search of the binned and averaged power
spectrum is
\begin{equation}
(P_{noise} > P) = 1 - (1 - Q(2MWP/P_{fit}| 2MW))^{n_{freq}}.
\end{equation}




\begin{footnotesize}

\end{footnotesize}

\clearpage






\end{document}